# Collective lattice excitations in the dynamic route for melting hydrodynamic 2D-crystals.

Mikheil Kharbedia[1], Niccolò Caselli[1,2], Macarena Calero[1,2,3], Lara H. Moleiro[1,2], Jesús F. Castillo[1,2], José A. Santiago[4,*], Diego Herráez-Aguilar[5,*] and Francisco Monroy[1,2,*]

[1]*Department of Physical Chemistry, Universidad Complutense de Madrid, Ciudad Universitaria s/n E28040 Madrid, Spain.*
[2]*Translational Biophysics, Instituto de Investigación Sanitaria Hospital Doce de Octubre, Av Andalucía s/n, E28041 Madrid, Spain.*
[3]*Faculty of Health Sciences - HM Hospitals, University Camilo José Cela, Urb. Villafranca del Castillo, E28692, Villanueva de la Cañada, Madrid, Spain.*
[4]*Departamento de Matemáticas Aplicadas y Sistemas, Universidad Autónoma Metropolitana Cuajimalpa, Vasco de Quiroga 4871, 05348 Ciudad de México, México.*
[5]*Instituto de Investigaciones Biosanitarias, Universidad Francisco de Vitoria, Ctra. Pozuelo-Majadahonda, E28223 Pozuelo de Alarcón, Spain.*

(Submitted on March 16[th], 2024)

Surface stiffnesses engender steady patterns of Faraday waves (FWs), so called hydrodynamic crystals as correspond to ordered wave lattices made of discrete subharmonics under monochromatic driving. Mastering rules are both inertia-imposed parametric resonance for frequency-halving together with rigidity-driven nonlinearity for wavefield self-focusing. They harness the discretization needed for coherent FW-packets to localize in space and time. Collective lattice excitations are observed as dispersionless propagating dislocations that lead periodic modulations arising from explicit symmetry breaking. In a field theory perspective, a halving genesis for the collective distorting modes is revealed as the natural pathway for hydrodynamic crystal melting.

Hydrodynamic crystals constitute an exotic status of oscillatory fluid matter appearing far from equilibrium as ordered wave lattices frozen under intrinsic interactions [1,2]. Intrinsically fluid but organized as stationary solidlike patterns, those structures provide a useful laboratory for testing non-equilibrium [3,4]. Hypothetically, lost fluid invariances should impart order as condensing stiffnesses in conservative flowing structures [5-7]. By following the concept at exploiting Faraday waves (FWs) [8], an inviscid macroscopic realization for a stable, free-standing crystal has been achieved in the shear-rigidized surface of water [9]. In general, the Faraday instability becomes frequency-halving as stems on the parametric resonance of the fluid surface under forcing the liquid against bulk inertia [8-10]. Hence, a steady and discretized assembly of nonlinear capillary ripples comprises emergent FW-packets [11], which constitute the analogue "atoms" of the novel hydrodynamic "crystal" in two dimensions [1,2,9]. Due to the "intrinsic" surface rigidness in a concert with capillary dispersion to imparting spatial symmetry [6,7,10,12], the coherent FW-packets became focused, or "frozen" into a square lattice with a locked wave phase [9]. Such systemic FW-condensation occurs not only in the absence of bulk stresses neither elastic [13], nor frictional [14], but also devoid of surface bounding constraints that could elicit spurious periodicities into the extrinsically guided FW-organization [15-17]. Yet, the Faraday interaction could eventually become the Achilles' heel for the hydrodynamic crystal; if newer subharmonic trajectories bifurcate from the FWs [18,19] wave disorder could then replicate into slower dynamical states associated to lattice defects [9,20]. Hence, an inverse cascade of disarranging modes could develop up to large scales, forcing lattice interactions to disentangle in a chaotic pathway towards crystal melting akin the turbulence in nonequilibrium systems [21,22].

In this work, we exploit the novel hydrodynamic crystals (HCs) to study the line defects created in the FW-lattice as the natural route for melting. Figure 1 portrays the hydrodynamic wavefield with a mastering role not only in lattice formation but also in its dislocation dynamics. Alike classical crystals, HCs primarily distort through of line dislocations occurred along the principal lattice axes [23]. Although lattice dynamics has been already studied in FW-patterns made of granular matter [21], or in surface bounded realizations under bulk frictional control [13,24], the collective melting instabilities have not been reported hitherto. Previous works assimilated the collective modes to elastic waves induced by an amplitude modulation retained under viscous drag [25,26]. An effective elasticity was invoked as a putative regulator [26]; however, no material substrate existed for surface wave ordering other than extrinsic bulk friction. In our ordered FW-lattice, we showed wavefield focusing under surface stiffening to be condition *sine qua non* for hydrodynamic crystallization [9]. We demonstrate field defocusing with a crucial role in lattice distortion under explicit symmetry breaking. Analogous to the Taylor's dislocations appeared in solids under mechanical stress (or heating) [23], lattice modes are observed to propagate as mobile line defects that interact each other as collective excitations forcing the crystal to disentangle (Suppl. Movie M1). A unified field perspective is adopted in a dynamic description for both the primary Faraday interaction responsible for crystal formation, and the secondary instabilities steering its excited, halving route towards turbulent melting [19].

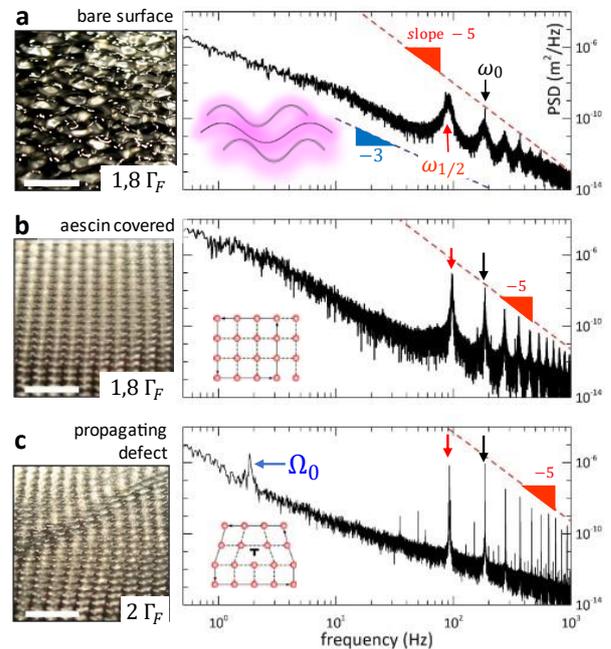

**Fig. 1. Hydrodynamic crystal and dislocation mode.** Discrete Faraday waves (FWs) in the capillary wave (CW) domain ($\omega_0 = 185\,Hz \gg \omega_{cap}$) at $\Gamma > \Gamma_F \approx 0.2$; (left panels; scale bars 5 mm). Wavefield spectral densities detected by LDV as discrete cascades of harmonics $\omega_n = n\omega_0$, of the fundamental mode $\omega_0$ (black arrow; right). They decay at inertial scaling ($PSD \equiv P_{\psi\psi} \propto \omega^{-5}$; red), including the Faraday mode ($\omega_{1/2} = \omega_0/2$; red arrow), and overtones ($\omega_{n/2} = n\omega_{1/2}$). Stochasticity is evident as a noisy background (friction-filtered $\sim\omega^{-3}$; blue). **a)** Disordered FWs in a bare water surface ($1.8\Gamma_F$). **b)** FW-lattice in an adsorbed layer of aescin (0.4mM); same conditions as in a). An organized FW-field emerges as a square-lattice for which the spectral peaks narrow. **c)** Lattice excitation as a propagating line defect (at $2\Gamma_F$); a structural peak emerges at $\Omega_0 = 1.8\,Hz \ll \omega_0$.

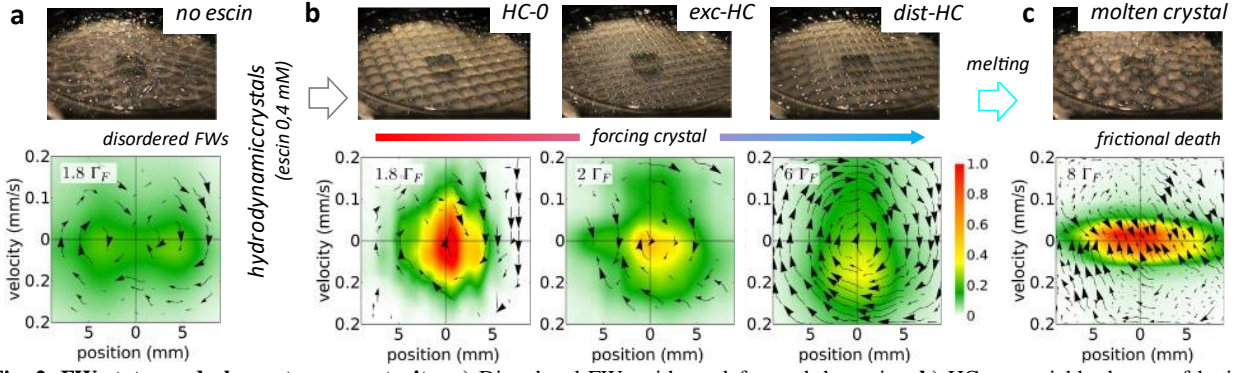

**Fig. 2. FW-states and phase stream portraits: a)** Disordered FWs with an defocused dynamics; **b)** HCs at variable degree of lattice distortion; *HC-0*) Zero-energy state; near-perfect crystal in a focused field promoting resonant orbits around the ground state; *exc-HC*) Excited state with nascent dislocations from field defocusing; *dist-HC*) Distorted lattice with collective excitations leading peripheral resonant dynamics. **c)** Molten crystal under frictional death.

*Experiment.* Discrete FWs do exist as frequency-halving resonances occurred when a fluid under gravity is subjected to vertical vibration above a critical acceleration; $a_F = 8\mu(\rho/\sigma)^{1/3}(2\pi\omega_0)^{5/3}$ in cylindrical vessels ($\omega_0$ is the driving frequency (see Suppl. Fig. S1); fluid characteristics: $\mu$ kinematic viscosity; $\rho$ density; $\sigma$ surface tension) [27]. The vessel is vibrated by a louder under monochromatic driving $A(t) = A_0 \cos(\omega_0 t)$ whereby one imposes wavefield amplitude $\psi_0$, and acceleration $a_0 \equiv A_0\omega_0^2$. As referred to the acceleration of gravity $g$, the reduced driving parameter $\Gamma \equiv a_0/g$ is varied in the Faraday interval; for water $\Gamma \geq \Gamma_F (\equiv a_F/g) \approx 0.2$. To realize HCs on highly tensioned water surfaces ($\sigma \approx 0.07\ N/m$, $a_F \approx 2\ m/s^2$), we exploit the same rationale as in Ref. [9], which yields regular squared FW-lattices of large aspect based on capillary waves (CWs; see Suppl. Fig. S2). By adding 0.4 mM Aescin, the decreased surface tension favors Faraday waving ($\sigma \approx 0.04\ N/m$), whereas the shear rigidity of the adsorbed monolayer ($G \approx 1\ N/m$) causes tough but reversible surface stiffening $\kappa \equiv G/\sigma \gg \kappa_{fr} \approx 2$, as phenomenologically determined [9]. Above the Faraday threshold ($\Gamma \geq \Gamma_{fr} \approx 1.8\Gamma_F$), we consider squared lattices made of capillary waves (CWs) excited at $\omega_0 = 185\ Hz \gg \omega_{cap} = (4\rho g^3/\sigma)^{1/4} \approx 15Hz$; here, the CWs resonate with the oscillating bulk flow giving rise to the distinctive FW-subharmonic at $\omega_{1/2} \equiv \omega_0/2$ [8-10], which fixes the master wavevector of the dispersionless wavefield $k_{1/2}^3 \equiv \rho\omega_{1/2}^2/\sigma$ (Suppl. Fig. S3) i.e., we consider a conservative wave lattice with a repetitive unit cell size ($\lambda_{1/2} \approx 1.8mm$). Neither boundary nor meniscus effects are relevant in this setup [9]. The surface wavefield $\psi(\mathbf{r},t)$ is analyzed by laser Doppler vibrometry (LDV), which measures the velocity trace $\dot\psi(t)$ detected in a single surface emplacement at phase lock-in with driving. Vertical displacements are obtained by stepwise integration $\psi(t_{i+1}) = \psi(t_i) + \dot\psi(t_i)\Delta t$ (being $\Delta t$ the time step); we compute the power spectral density as $PSD(\omega) \equiv P_{\psi\psi} \equiv \int|\psi(t)|^2 e^{-i\omega t}dt = P_{\dot\psi\dot\psi}/\omega^2$ (Fig. 1). Because natural stochasticity, the phase portraits are charted as densities of states (PDFs) explored along the time series; the statistical energy landscape $U \equiv -lnPDF$. Trajectory circulations are calculated as probability currents in phase space.

*Hydrodynamic crystal freezing.* Figure 1 depicts the crystal freezing scenario. In the absence of aescin, the FWs appear completely disordered (at $\Gamma = 1.8\Gamma_F$; Fig. 1a, left). The shared LDV-spectral features are (right panels): i) Ordinary cascade of discrete harmonics $\psi^{(0)}\{\omega_n = n\omega_0\}$ decaying as $P_{\psi\psi} \sim \omega_n^{-5}$ (*o*-field at inertial scaling); ii) Extraordinary subharmonic cascade $\psi^{(1/2)}$, constituted by leading half-harmonic $\omega_{1/2}$, and overtones $\omega_{n/2} = n\omega_{1/2}$ (*e*-field at parametric resonance). Indeed, the nonlinear FW-field is composed by the superposition of both cascades ($\psi \equiv o \cup e$), which appear relatively broaden in the fluid case (Fig. 1a; right). At $1.8\Gamma_F$-forcing, the presence of aescin "freezes" the FWs into HCs as regular square-lattices of large aspect under wavefield self-focusing (Fig. 1b; left); FW- spectral features are preserved in such focused state, whereas evident peak narrowing occurs at increasing crystal order (Fig. 1b; right) [9]. Under further weak energy increase ($2\Gamma_F$), however, a single dislocation appears (Taylor-like), causing a mobile lattice distortion (Fig. 1c, left). This excitation is spectrally identified as a propagating mode ($\Omega_0 \approx 1.8 Hz \ll \omega_{1/2}$; Fig. 1c; right).

Figure 2 shows a sequence of steady FW-states giving rise to HCs with a melting dynamics for increasing energy. A nontrivial evolution is identified in the systemic circulations monitored in phase space. Whereas the disordered FWs behave as a delocalized oscillator (Fig. 2a), the ordered HCs are characterized by a centripetal dynamics typical for a locked resonator (Fig. 2b; HC-0 ground state) [28]. Further HC-distortion causes lattice dynamics to evolve from a slight centrifugal trending, through of more external orbits in a weakly defocused field (slightly excited crystal; *exc*-HC at $2\Gamma_F$), up to evident peripherical orbiting under field defocusing (distorted lattice at $6\Gamma_F$; *dist*-HC). Finally, a molten state is undergone with a collapsed dynamics dominated by frictional penalties to fast motions (at $\Gamma \geq 8\Gamma_F$; Fig. 2c); here, the largest displacements are favored by the slowest velocities. As being cornerstone for the present work, the lattice dynamics is analyzed below.

*Propagating defects and collective distortions.* Figure 3 shows LDV-spectra measured at four stages of lattice distortion. The single dislocations appear only at small deformation ($\Gamma \approx 2\Gamma_F \gtrsim \Gamma_{fr}$; Fig. 3a); their propagative characteristics evidence a material elasticity enabled by in-plane stiffness. As mastered by the rigidity modulus, the natural frequency is fixed as $\Omega_0 \cong (G/\rho_s)^{1/2}$ (being $\rho_s$ a surface equivalent mass density). Because $\Omega_0 \approx 1.8 Hz$, one estimates $\rho_s \approx 0.3 kg/m^2$ (taking $G = 1 N/m$), which represents an inertial layer of thickness $h = \rho_s/\rho \approx 1\ mm$ (on the same order than wave dimensions $h \approx \psi_0 \geq \lambda_{1/2}$). At $4\Gamma_F$, we detect the fundamental peak $\Omega_0$ followed by a complete cascade of harmonics ($2\Omega_0, 3\Omega_0$ ...), with a scaling structure resembling the Faraday resonance ($\omega^{-5}$-inertial decay over an $\omega^{-3}$-frictional background; Fig. 3b, left). The single dislocations self-organize as a collective lattice distortion of wavelength $\Lambda \approx 3.2\ cm$ ($Q \equiv 2\pi/\Lambda \approx 2\ cm^{-1}$), actually a spatial modulation $\Lambda = N\lambda_{1/2}$ spanning $N \approx 18$ unit cells (Fig. 3; right). Assuming crystal homogeneity, we estimate the collective mode propagating at constant velocity $c = \Omega_0/Q \approx 1\ cm/s$, in qualitative agreement with visual observation (Suppl. Movie M2). Additional forcing at $6\Gamma_F$ causes cascade excitations with a canonical Faraday structure composed by ordinary harmonics $\Omega_n = n\Omega_0$, and extraordinary subharmonics $\Omega_{n/2} = n\Omega_{1/2}$, which emerge from the half-frequency structural mode; this is $\Omega_{1/2} = \Omega_0/2 \approx 0.9 Hz$ (Fig. 3c; left). Such strong lattice distortions appear as collective defect modulations (Fig. 3c; right). Further forcing at $8\Gamma_F$ makes the structural band to superpose as a Landau's continuum compatible with the frequency-halving instability behind fluid turbulence [18,19]. Peak broadening, mode superposition, and vanishing inertial scaling in favor of frictional decay are characteristic for this "molten" state (Fig. 3d; left), which waves steadily turbulent (right image). These behaviors are systematically reproduced at different $\omega_0$ and $\Gamma$, leading lattice defects with a propagating dynamics underneath. On the one hand, we detect the structural mode to be nondispersive, i.e. $\Omega_0 = c_\Omega Q$, with a constant propagation velocity as $c_\Omega = 1.1 \pm 0.1\ cm/s$ (Fig. 3e), in agreement with above estimations. On the other hand, dissipation holds at high forcing; indeed, the spectral peaks broaden under $\Gamma$-induced unsteadiness (Fig. 3f). The most ordered crystal is identified by the sharpest spectral peaks (at $\Gamma_{fr} \approx 1.8\Gamma_F$), which become broader across a continuous melting; the chaotic melt shows similar unsteadiness as the bare water surface (Fig. 3f). By expanding on the Landau's conjecture for chaotic flows [18,29], we argue the observed unsteadiness as essential for HC-melting via lattice breakdown (Suppl. Movie M3).

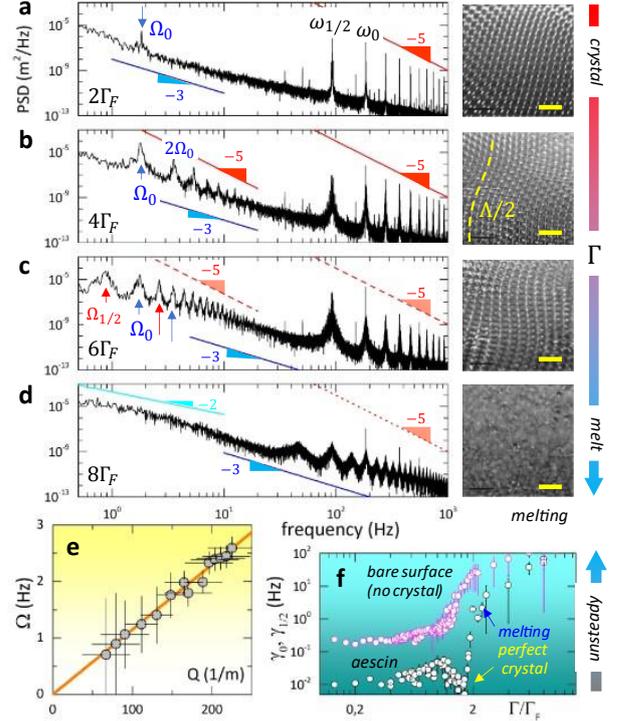

**Fig. 3. Crystal melting under collective lattice excitation.** HCs distorted upon increasing forcing ($\Gamma$): *Left)* LDV-spectra. The FW-field appears as a nonlinear cascade from the natural frequency ($\omega_0 = 185$ Hz). The (sub)harmonic decay is governed by inertia ($\sim\omega^{-5}$). The spectral background is limited by friction ($\sim\omega^{-3}$). Straight lines represent expected scaling; dashed lines are an eye-guide for vanishing behavior. Structural distortions do appear in the low-$\omega$ band. *Right)* Real lattices (scale bar 5 mm). **a)** Weakly distorted square lattice at $2\Gamma_F$, with single dislocations propagating at $\Omega_0 \approx 1.8 Hz$. **b)** Collective lattice modes at $4\Gamma_F$; composed by an inertial *o*-cascade ($\sim\Omega^{-5}$), of structural modes (fundamental $\Omega_0$, and higher harmonics $2\Omega_0, 3\Omega_0, ...$ *left*), causing collective lattice distortion of wavelength $\Lambda$ (*right*). **c)** Distorted lattice at $6\Gamma_F$. The collective mode undergoes frequency-halving at $\Omega_{1/2} \approx 0.8\ Hz \approx \Omega_0/2$, with *e*-cascade at $n\Omega_{1/2}$, superposed to *o*-cascade at $n\Omega_0$. **d)** Molten crystal at $8\Gamma_F$. Low-frequency distortions inflate superposed under frequency-halving in a Landau's continuous ($\sim\omega^{-2}$). **e)** Dispersion relation for the structural mode. *Symbols)* experimental data; *straight line)* linear fit to nondispersive relationship. **f)** $\Gamma$-dependence of spectral unsteadiness measured as peak widths for: fundamental mode ($\gamma_0$); Faraday subharmonic ($\gamma_{1/2}$). *Experimental series:* bare surface (magenta); HC (white).

*Potential energy: field theory.* Figure 4a shows the energy landscapes calculated from the experimental PDFs for the reduced displacements $\psi \to \psi/\psi_0$. Different interactions and symmetries are revealed in representative scenes. From a field theory perspective accounting for symmetries [30], the potential energy expands as:

$$U(\psi)/U_0 = k\psi^2/2 + \alpha\psi^3/3 + \beta\psi^4/4 + \delta\psi^6/6 \quad (1)$$

in terms of dimensionless spring constant ($k$), and anharmonic couplings ($\alpha, \beta$ and $\delta$). The zeroth energy

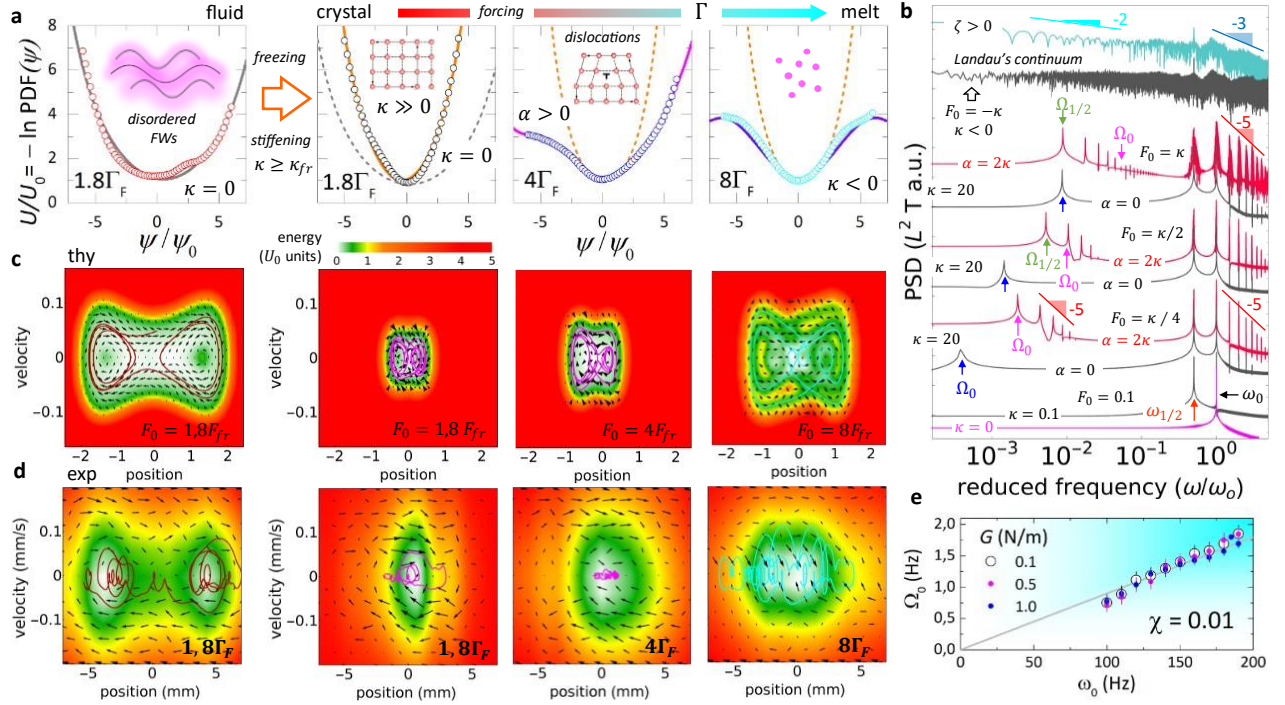

**Fig. 4. Energy landscapes and NLO dynamics. a)** Potential energy in representative states (experimentally inferred as $U/U_0 = -lnPDF(\bar{\psi})$ for $\bar{\psi} = \psi/\psi_0$; normalization $U_0 = -lnPDF_{max}$ and $\psi_0 = 2mm$). Symbols: experimental data. Straight lines: best fits to the NLO-potential in Eq. 1 (Disordered FWs $k = -0.1$, $\kappa = 1.1$, $\alpha = 0$; near-perfect crystal $k = 1$, $\kappa = 19.8$, $\alpha = 0$; distorted crystal $k = 1$, $\kappa = 19.9$, $\alpha = 40.3$; molten crystal $k = 0.9$, $\kappa = -3.2$, $\alpha = 0$; 10% typical uncertainty; reduced units referred to $m = 1$). All NLO-potentials are corralled by a perimetric wall as $U = U_{NLO} + \delta\bar{\psi}^6$ ($\delta < 1$), which accounts for effective surface confinement ($U \approx U_{NLO}$ for $|\bar{\psi}| < 1$ and $U \to +\infty$ for $|\bar{\psi}| \gg 1$). **b)** NLO-spectra at increasing force $F_0$ (NLO-parameters inferred from a). Main parametric resonance features are predicted including Faraday field $\psi^{(0)} \cup \psi^{(1/2)}$, and modulation modes $\Omega_n$. **c)** Theoretical phase portraits for the NLO-resonator at representative states (same potentials as in a); forcing amplitude $F_0$ by reference to $F_{fr} = k\psi_0$. A representative trajectory during a $10^3/\omega_0$-cycle is snapshot over the calculated energy landscape. Arrows indicate global circulations in phase space. **d)** Experimental phase portraits for the corresponding FW-status (same symbols as in c). **e)** Experimental lattice susceptibility for the HC-collective modes at variable surface stiffening. Straight line: best fit to a linear dependence for amplitude modulation modes.

$U_0 \equiv k\psi_0^2/2$ is chosen to determine the normalization constant at forcing amplitude $F_0 = k\psi_0$ i.e., $U_0 = F_0\psi_0/2$. Timescales are referred to $\omega_0$, so that $k \equiv \sigma_{eff}/m\omega_0^2$ and $\beta \equiv G_{eff}/m\omega_0^2$, in terms of effective stiffnesses and inertial mass $m \equiv F_0/A_0\omega_0^2$ (Suppl. Note N1) [9]. Because rotational invariance (fluidity), the vacuum state enables parity symmetry being constituted by harmonic response $k > 0$, plus quartic coupling $\beta > 0$ [10]; it encodes the Faraday interaction [9,10], as well the nonlinear (three-wave) generator under field self-interaction $\psi^2 \mapsto \varphi$ [34]. False vacuum states account for effective fluid metastability ($k < 0$) [31]. The quartic structure translates the square-lattice $C_4$-symmetry into an equivalent harmonic FW-field; if $\varphi = \sqrt{2}\psi^2/2$, we cast $U_\beta = \beta\psi^4/4 \mapsto k_\beta\varphi^2/2$ with quartic spring constant $k_\beta \equiv \beta$ [9,30]. The stiffening parameter is given as $\kappa \equiv \beta/k \Rightarrow G/\sigma$. Parity symmetry is lost under cubic coupling ($\alpha \neq 0$). In order to stable vacuum a bounding term is considered ($0 < \delta < |\beta|$), enabling surface integrity as a mean-field confinement.

The disordered FW-liquid is governed by slightly repulsive harmonic response ($k \lesssim 0$), under marginal quartic nonlinearity $U \approx U_L \approx k\psi^2/2(1 + \kappa\psi^2/2)$ (at $\kappa < \kappa_{fr}$; Fig. 4a; left). At $1.8\Gamma_F \approx \Gamma_{fr}$, crystal FW-freezing occurs as field self-focusing under relevant stiffening ($\kappa \gg$ 1). Above a freezing point $\kappa > \kappa_{fr} = 2$, at enough driving force $F_0 > F_{fr} = 3(2k^3/\beta)^{1/2}$ (at $U_\beta > U_L$), the ground HC-state holds $U_{HC}^{(0)} \approx U_L + \beta\varphi^2/2$; for the real crystal, we estimate $\kappa \approx 20 \gg \kappa_{fr}$ (Fig. 4a; second panel). At $4\Gamma_F$, collective lattice distortion happens under explicit symmetry breaking (Fig. 4a; third panel). Here, parity is broken as a cubic excitation $U_{HC}^{(exc)} \approx U_{HC}^{(0)} + \alpha\psi^3/3$ (for $\alpha^2 \geq 4k\beta \gg 0$). It represents the structural mode $\Omega_0$ (and relatives $\Omega_n$) shifting towards $\psi_0^{(exc)} \approx -\alpha/\beta$ in an effectively rigidified spring $k^{(exc)} \approx k + \alpha^2/\beta$, at enhanced anharmonicity $\beta^{(exc)} = \beta + 2\alpha/3$ [30]. At $8\Gamma_F$, the field evolves into a completely defocused state affected by a repulsive quartic interaction ($\beta \ll 0$; Fig. 4a, right); here, the molten crystal becomes stable ($U_{melt} < U_{HC}$), although potentially unsteady (Fig. 3f). These fields are further tested via dynamical analysis.

*Rigidity-harnessed dynamics in a parametric resonator.* Inspired on the Kolmogorov-Zakharov (KZ) theory for the nonlinear capillary fabric [32,33], the surface stiffness has been argued to constitute the dynamic skeleton for FW-condensation [9]. The KZ-field fixes the spatial structure imposed by Laplace stress, and a wave colliding kinetics encoded in the nonlinear Schrödinger (NLS) equation [34]:

$$i\psi_t + k\nabla^2\psi + \beta|\psi|^2\psi = 0, \qquad (2)$$

which describes nonlinear, inviscid propagation under CW-genesis by three-wave coupling [18]. However, NLS dynamics is devoid of the momentum ingredient necessary to engender the FW-field under parametric resonance. Assuming NLS essentials at compatibility with the Stokes flow involved in FW-excitation (see Suppl. Note N1), we postulate a minimal but comprehensive nonlinear oscillator (NLO) with a forced field structure inherited from Eq. (1); under homogenous forcing and viscous dissipation, hence we restate the wavefield equation:

$$m\psi_{tt} + \zeta\psi_t + k\psi + (\alpha + \beta\psi + \delta\psi^3)\varphi = F(t) \qquad (3)$$

Such NLO-generator enables Faraday (CW-) resonances between wave inertia (as represented by the inertial mass m), and the monochromatic source $F(t) = F_0 \cos(\omega_0 t)$. It drives a spatially homogeneous, time-discretized response $\psi = \psi^{(0)} + \epsilon\psi^{(1/2)}$ composed by main field $\psi^{(0)}$ lead by the natural response at $\omega_0 = (k/m)^{1/2}$, and subsidiary ($\epsilon < 1$), subharmonic field $\psi^{(1/2)} \propto \psi^2 \mapsto \varphi$ bifurcating under parametric resonance at self-interaction, this is $\ddot\varphi = -\omega_{1/2}^2(\psi + \alpha/\beta)\varphi$ [5,19]. The resonant half-harmonic $\omega_{1/2}^2 = \beta/m_{1/2}$ arises from quartic coupling ($\beta > 0$), which generates FW-packets of mass $m_{1/2}$. Since $\omega_{1/2} \equiv \omega_0/2$, thus $m_{1/2} \equiv 4mk$ i.e., field stiffening elicits denser $\varphi$-inertia. The kinetic term assimilates to a generalized drag characterized by a complex friction coefficient $\zeta \equiv \zeta_{visc} + i\omega_0 m$; in the inviscid case ($\zeta_{visc} = 0$), it reduces to the inertial NLS structure ($\zeta = i\omega_0 m$). This elemental NLO-dynamics was shown in accordance with phenomenology [9]; particularly, inertia-driven $\omega^{-5}$-cascades, and $\omega^{-3}$ frictional background. Generalized frequency- and phase-locking, frequency-halving and systemic discretization were also forecasted as NLO-FW distinctives allowing the Faraday bifurcation that builds the hydrodynamic crystal. Because ruling collective excitations under Faraday instability, we look at broken symmetries with a crucial role in making crystal melting to happen.

Figure 4b shows NLO-predictions in agreement with experimental spectra (Fig. 3). At $\kappa \gg \kappa_{fr}$, the parent cascade $\psi^{(0)}$, as well the FW-field $\psi^{(1/2)}$, appear both as HC-field generators (Fig. 4b; lower panels). Above the freezing point ($F_0 > F_{fr}$), a structural excitation appears isolated as an NLS-modulation instability [35,36]; this emerges from the FW-field as a non-dispersive, massive mode of frequency $\Omega_0 \propto \omega_0$ (Suppl. Note N3). In accordance with the cubic asymmetry observed in distorted lattices (Fig. 3b), a structural $\Omega_n$-cascade appears weather parity is explicitly broken ($\alpha > \beta \gg 0$); since effective stiffening takes place ($\beta^{(exc)} > \beta$), secondary resonances refocus within this excited state. Hence, subsequent NLO-forcing causes $\Omega_{n/2}$-halving from such an $\alpha$-perturbed state in a similar way to collective distortions in the real lattices (Fig. 3c). At high forcing, Landau's continua are forecast in defocusing potentials ($\kappa < 0$); they represent a symmetry lost across the melting transition [37]. NLO-friction also limits unsteadiness, particularly at high energy (Fig. 4b, upper). Figure 4c shows theoretical phase-portraits in their own energy landscapes as NLO- predictors at compatibility with peripheral orbiting observed in real lattices (Fig. 4d). Even in the presence of noise the trajectories become focused under lattice freezing, whereas they experience defocusing under symmetry distortion (Suppl. Movies M4-M7). These numerics reveal quartic self-interactions as key governors for the Faraday bifurcating organization in the dynamic pathway from the crystal to the molten fluid.

Collective lattice distortion, and eventual melting are shown to occur as an explicit symmetry breaking generated from the ground crystal state upon collective Faraday excitation (Fig. 4c-d). Their inertial massive nature has been unequivocally shown as far the dislocation modes propagate dispersionless (Fig. 3e). They appear as a modulation mode under inertial control [9]; being susceptible to emerge at frequency $\Omega_0 = \chi\omega_0$, it is determined by the cell size (as fixed by $\omega_0$, through of the lattice susceptibility $\chi$). Figure 4d shows experimental data for different shear rigidities spanning a broad range above the freezing point ($G \geq G_{fr} \approx 0.1 \, N/m$). The experimental results share a common susceptibility $\chi = 0.010 \pm 0.006$, independently of surface freezing and external forcing. From a matter wave perspective, the pilot FW-packets locate in the nodes of the lattice as pseudo-particles with inertial mass $m_{1/2}$ concentrated over their wavelength $\lambda_{1/2}$ [9,11]. The collective distortions stem congruently on the lattice "atoms" (the FW-packets), appearing as massive excitations explicitly breaking lattice symmetry (density waves as pseudo-Goldstone lattice modes) [24,38]; they span a collective wavelength $\Lambda = N\lambda_{1/2}$, and comprise a total mass $m_G = N^2 m_{1/2}$, thus involving inertial resonance under surface density $\rho_s \equiv m_G/\Lambda^2 \equiv m_{1/2}/\lambda_{1/2}^2$. Because propagating nature $\Omega_0^2 = (G/\rho_s)Q^2$, since $m_{1/2} = 4mG/\sigma$, then $\Omega_0 = \pi\omega_0/N^2$, we deduce a susceptibility $\chi = \pi/N^2$; akin classical crystals, it solely arises from defect size and lattice symmetry but not depends on stiffness [23]. For observed $N \approx 18$ (Fig. 3), one estimates $\chi \approx 0.01$ in agreement with experiments (Fig. 4c).

As a first principle, the ground crystal state emerges from the Faraday interaction engendered upon surface stiffening. In a concert with wave inertia, it harnesses the discrete resonances needed to crystallize the FWs into coherent packets, i.e. pilot waves localized in a regular lattice. Our study reveals the crystal defects as slow lattice modulations arising from bifurcated FW-resonances. Because energy is finite ranged under these excitations, lattice symmetry is explicitly broken by the collective distortions associated with the local reorientation of the parity symmetry picking out one direction in the crystal. Such massive modes become a collective excitation emerged among the discrete, coherent, and momentum-containing FW-packets, i.e. the "atoms" that constitute the hydrodynamic crystal.

To summarize, a genuine macroscopic 2D-realization has been delivered for distorted hydrodynamic crystal lattices entailed by frequency-halving Faraday interaction. Above the freezing point, lattice order is progressively lost by means of disorganizing dislocations that propagate under explicit symmetry breaking, just alike lattice vibrations in classical crystals. Because collective distortions emerge from the parent FW-lattice, they dominate the dynamic route towards crystal melting with a ubiquitous (Faraday) hierarchy. In a systemic inflation from the distorted crystal up to a chaotic melt, these secondary instabilities progressively halve in a disordering mode continuum akin the Landau's picture of nascent turbulence. As a new insight, we present a dynamic paradigm for the collective 2D-lattice excitations as the natural route for hydrodynamic crystal melting. The adopted field perspective settles a macroscopic laboratory for testing inflationary dynamics under systemic halving bifurcation.

This work was supported by Spanish funding agencies: Ministerio de Ciencia, Innovación y Universidades – Agencia Estatal de Investigación (AEI) under Grants No. PID2019-108391RB-I00 and TED2021-132296B-C52, and Comunidad de Madrid under Grants No. Y2018/BIO-5207 and S2018/NMT-4389.

*Correspondence authors:*
José A. Santiago. jasantiagog@gmail.com
Diego Herráez-Aguilar. diego.herraez@ufv.es
Francisco Monroy. monroy@ucm.es